\documentclass[conference]{IEEEtran}

\usepackage{graphicx}
\usepackage{latexsym}
\usepackage{amsfonts}
\usepackage{amssymb}
\usepackage{amsbsy}
\usepackage{amsmath}
\usepackage{multicol}
\usepackage{float}
\usepackage{subfigure}
\usepackage{algorithm}
\usepackage[dvips]{epsfig}

\usepackage{fancyhdr}
\usepackage{epsfig}
\usepackage{threeparttable}
\usepackage{epsf,epsfig}
\usepackage{amsthm}
\usepackage[noadjust]{cite}
\usepackage{dsfont}
\usepackage{color}
\usepackage{enumerate}
\usepackage{setspace}

\usepackage{amsmath}
\usepackage{algorithm}
\usepackage[noend]{algpseudocode}
\algrenewcommand{\algorithmiccomment}[1]{\hskip3em$\triangleright$ #1}

\makeatletter
\def\BState{\State\hskip-\ALG@thistlm}
\makeatother

\newtheorem{theorem}{Theorem}
\newtheorem{remark}{Remark}
\newtheorem{lemma}{Lemma}
\newtheorem{proposition}{Proposition}
\newtheorem{assumption}{Assumption}

\newtheorem{definition}[theorem]{Definition}

\def\SIR{\mathsf{SIR}}

\def\l{\left}
\def\r{\right}
\def\({\left(}
\def\){\right)}
\def\[{\left[}
\def\]{\right]}

\def\nn{\nonumber}

\usepackage{dblfloatfix} 

\title{\LARGE Sense-and-Predict: Opportunistic MAC Based on \\ Spatial Interference Correlation for Cognitive Radio Networks}
\begin{document}


\vskip -5pt
\author{\IEEEauthorblockN{Jeemin Kim, Seung-Woo Ko, Han Cha and Seong-Lyun Kim}
 \IEEEauthorblockA{School of Electrical and Electronic Engineering, Yonsei University, S. Korea}
  \IEEEauthorblockA{Email: \{jmkim, swko, chan, slkim\}@ramo.yonsei.ac.kr
}
}

\maketitle
\begin{abstract}
Opportunity detection at secondary transmitters (TXs) is  a key technique enabling cognitive radio (CR) networks. 
Such detection however cannot guarantee reliable communication at secondary receivers (RXs), 
especially when their association distance is long. 
To cope with the issue, this paper proposes a novel MAC called  sense-and-predict (SaP), where each secondary TX decides whether to access or not based on the prediction of the interference level at RX. Firstly, we provide the spatial interference correlation in a probabilistic form using stochastic geometry, and utilize it to maximize the area spectral efficiency (ASE) for secondary networks while guaranteeing the service quality of primary networks. 
Through simulations and testbed experiments using USRP, SaP is shown to always achieve ASE improvement compared with the conventional TX based sensing.

\end{abstract}


\IEEEpeerreviewmaketitle

\vskip -10pt
\section{Introduction}

A cognitive radio (CR) has been considered as a promising technique to improve the efficiency of wireless spectrum 
by allowing the access of secondary users if the spectrum is under-utilized \cite{Zhao07}.
In general, a secondary transmitter (TX) senses the medium before accessing the spectrum to check the level of its utilization. 
Such a  sensing mechanism triggered by a transmitter, however, has two major bottlenecks.
First of all, it cannot detect the existence of primary receivers (RXs), making it difficult to protect the primary network.
Next, the interference level at the secondary TX differs from that at its RX where the transmitted signal actually encounters.  

\begin{figure}
\centering 
{\includegraphics[width=7cm]{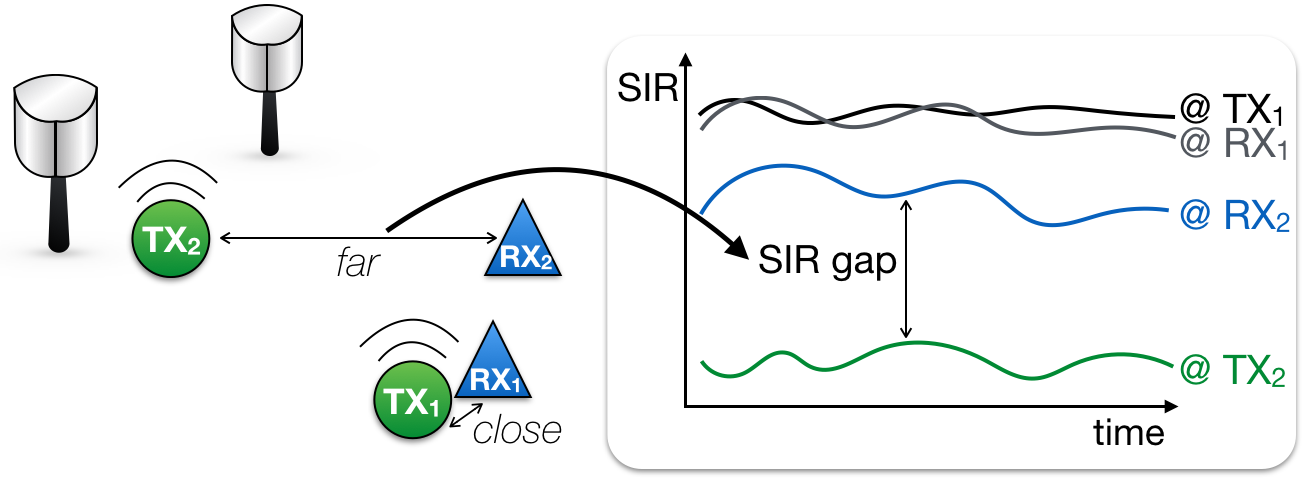}} 
\vskip -10pt
\caption{Illustration of the access opportunity detection problem due to secondary TX-RX distance separation. When secondary TX-RX association distance is long, the $\SIR$ measured at TX significantly differs from the $\SIR$ at RX (see the $\SIR$ gap between $\textsf{TX}_1$ and $\textsf{TX}_2$), leading to detection errors.\label{system}}
\vskip -15pt 
\end{figure}

There have been relatively many works to predict the locations of primary RXs as remedies of the first bottleneck \cite{Ben05, lan08} but the difference of interference levels of secondary TX and RX has still remained under-explored.
This is mainly because the distance between them has been regarded to be short enough to make the interference gap negligibly small~\cite{Mohammad15}.
In upcoming CR networks, on the other hand, this distance can become longer 
i) when a cellular system becomes the secondary network (e.g., license assisted access \cite{Ericsson15}, public safety long-term evolution \cite{Ferrus13}) and 
ii) when the secondary TXs need to use dedicated spectrum sensors due to their low device form factor \cite{Deng12}.
This tendency brings about larger interference gap between the secondary TX and RX, increasing errors in detecting the spectrum access opportunities as shown in Fig.~\ref{system}.

In view of tackling this issue, 
we propose a novel MAC design called {\it Sense-and-Predict} (SaP)
in which each secondary TX opportunistically accesses the spectrum 
by predicting the level of interference at its RX.
Specifically, given the measured interference, a secondary TX computes the probability 
that the level of interference at its RX is less than the predetermined threshold, 
and decides whether to access or not with this probability. 
The interference predictability is closely related to the correlation of interference between two locations. When they are co-located, it is obvious that they meet the same level of interference. As the distance between the two becomes longer, the similarity decreases and finally they are independent. We figure out this spatial correlation in form of the conditional probability with the aid of stochastic geometry (SG).  

It is worth mentioning that such a probabilistic approach based on SG plays a role to transform the real interference value to the level of the spectrum access opportunity in a stochastic manner. 
The measured interference level at TX per se cannot provide an accurate prediction of interference level at RX 
because it depends on various parameters, such as the density of TXs, transmit power et al.
For instance, even though the measured interference at TX seems to be relatively high, there can be more access opportunities at RX especially when the density of TXs is low. 
SG enables to reflect these kinds of parameters within one metric in a probabilistic form, which thus harmonizes with the operation of the proposed SaP.


Within the SaP framework, we define an area spectral efficiency (ASE) of the secondary network as 
the sum of throughput for secondary TXs within a unit area,
depending on the following two parameters.  
One is the access threshold to decide the access probability that we mentioned above.
It affects the density of the concurrent secondary transmissions in the network, 
determining the level of the aggregate interference. 
The other is the minimum signal-to-interference ratio ($\SIR$) required for decoding the transmitted secondary signal,
corresponding to the quality of each wireless link. 
Note that there exist important design issues in which these two parameters are involved. 
For example, a small access threshold encourages the secondary TX to access the medium more frequently 
but reduces the transmission success probability due to higher aggregate interference. 
To compensate the loss, we can decrease the minimum $\SIR$ but brings about the loss of each link quality. 
Reflecting on these relations, we jointly optimize the two parameters to maximize secondary network ASE while guaranteeing the protection of the primary network.   

\textbf{Related Works} -- 
Interference at neighboring nodes is sometimes measured analogously  because their topological environments like  interferers, reflectors, and scatterers are similar. Traditionally, the spatial interference correlation has been interpreted as an obstacle required to be overcome, 
such as the reduction of degree-of-freedom in   
multiple-input multiple-output (MIMO) communications when the antenna spacing is less than the half of wavelength \cite{mimo}. 
On the other hand, there are some trials to rather exploit the correlation for the purpose of enhancng the network performance.  
In \cite{Ali11} and \cite{Mal12}, for example, the spatial correlation is utilized to control the mobility of distributed robots. 
Specifically, by predicting the interference level at a spatially separated location, each robot can reduce the number of channel measurements to find the adequate location guaranteeing sufficient throughput.

The spatial interference correlation is in particular studied to improve the throughput of ad-hoc networks, where each node transmits in a distributed manner according to the local situation \cite{DM}, \cite{Ke07}.
In \cite{DM}, a spatially adaptive random access scheme is proposed, in which the transmitters in the ad-hoc network have different transmit probabilities due to the regional differences.
The proposed scheme is validated to achieve better performance than the conventional ALOHA scheme where all TXs access the spectrum with the same probability.
In addition, the authors in \cite{Ke07} take into account the spatial correlation between desired signal and interference in MIMO ad-hoc networks. They design a MAC protocol allowing concurrent transmissions most of which can be cancelled out due to the interference correlation.

SG has been understood as an efficient tool to offer the tractability in modeling interference at a typical point in wireless networks \cite{Andrews11}.
However, quantifying the spatial interference correlation with SG is far more challenging since the joint analysis of two spatial observations are required \cite{Haenggi13, Ganti09, Shankar16}. 
The authors in \cite{Haenggi13} and \cite{Ganti09} focus on the spatial interference correlation to analyze the link outage in ad-hoc network with SG.
By extending these works, the authors in \cite{Shankar16} derive the joint coverage probability at two spatial locations in a cellular network where mobile devices are moving. 
However, neither of them focus on the spatial interference dependency between secondary TX and RX in CR networks where the network analysis is more complicated due to the heterogeneity of primary and secondary networks. 
One of the goals of this paper thus is to fill this gap and provide useful insights for MAC algorithm in CR networks.

\textbf{Contributions and Organization} -- This paper proposes the SaP in which given the channel sensing result, the access probability of each secondary TX is fine-tuned based on the \emph{predicted} level of interference at its paired secondary RX. 
The main contributions are listed below.
\begin{itemize}
\item By using SG, we derive the access probability, a probability that the received $\SIR$ at a typical secondary RX exceeds a threshold under the condition of channel measurement at its paired TX, and its tight lower bound.
The analytic result indicates that the access probability becomes totally independent of the interference measurement if the distance between secondary TX and RX is far enough.
Through the MATLAB simulation and testbed experiment utilizing universal software radio peripheral (USRP), we verify the access probability derivation and demonstrate that significant improvements are feasible.
\item The design guideline for the optimal access threshold and target $\SIR$ for decoding is provided  to maximize the ASE while protecting the primary network. Along with the secondary TX density increment, the optimal access threshold increases whereas the optimal target $\SIR$ remains constant. On the other hand, both of the optimal threshold and target $\SIR$ increase with the secondary transmit power $P_2$. In an asymptotic case where the aggregate interference from the primary TXs is much higher than that from the secondary TXs, the optimal access threshold and target $\SIR$ are given in closed forms.
\end{itemize}

The rest of this paper is organized as follows.
Our system model and metrics are presented in Section II.
The access probability is derived in Section III, with the verification through the simulations and testbed experiments.
This result is utilized to maximize the ASE of secondary network while protecting the primary network in Section IV.
In addition, some realistic scenarios are briefly discussed in Section V. 
Finally, the concluding remarks are described in Section VI followed by the proofs of lemma and propositions in Appendix.

\vskip -5pt
\section{Models and Metrics}

\subsection{Network Model}
Consider a cognitive radio network where subscript $k\in\{1,2\}$ denotes primary and secondary networks, respectively. 
The coordinates of $k$-th network TXs follow a homogeneous Poisson point process (PPP) $\Phi_k$ with density $\lambda_k$. 
Processes $\Phi_1$ and $\Phi_2$ are independent of each other. 
Each secondary TX has its paired RX with a distance of $d$, and
the density of secondary RXs  is thus identical to  $\lambda_2$.

\vskip -10pt
\subsection{Channel Model}
The  $k$-th network TX sends a signal with power $P_k$. 
The transmitted signal experiences path-loss attenuation with the exponent $\alpha >2$ as well as Rayleigh fading with unity mean, i.e. $h \sim \exp(1)$. 
The primary and secondary networks share the same spectrum, resulting in co-channel interference.
By the aid of Slyvnak's theorem \cite{kendall}, the $\SIR$ at a typical RX of $k$-th network located at the origin is represented as

\small\begin{align}\label{Eq:SIR_Definition}
\SIR_k=\frac{P_k  h^{(0)} {x^{(0)}}^{-\alpha}}{\sum_{j=1}^{2}{\sum_{i \in \Phi_j} {P_j \phi_j^{(i)}  h^{(i)} {x_j^{(i)}}^{-\alpha} }}},
\end{align}\normalsize
where $x_k^{(0)}$ and $x_j^{(i)}$ respectively denote the distances to the associated TX and the $i$-th nearest interfering TX of the $j$-th network from the typical RX. 
The value ${\phi}_j^{(i)}$ denotes a probability that the $i$-th nearest TX in the $j$-th network transmits a signal. 
We assume that every primary TX keeps on transmitting a signal, i.e. $\phi_1^{(i)}=1, \forall i$, 
whereas a secondary TX computes its own $\phi_2$ 
depending on its MAC design. 

\begin{figure}
\centering 
{\includegraphics[width=6.5cm]{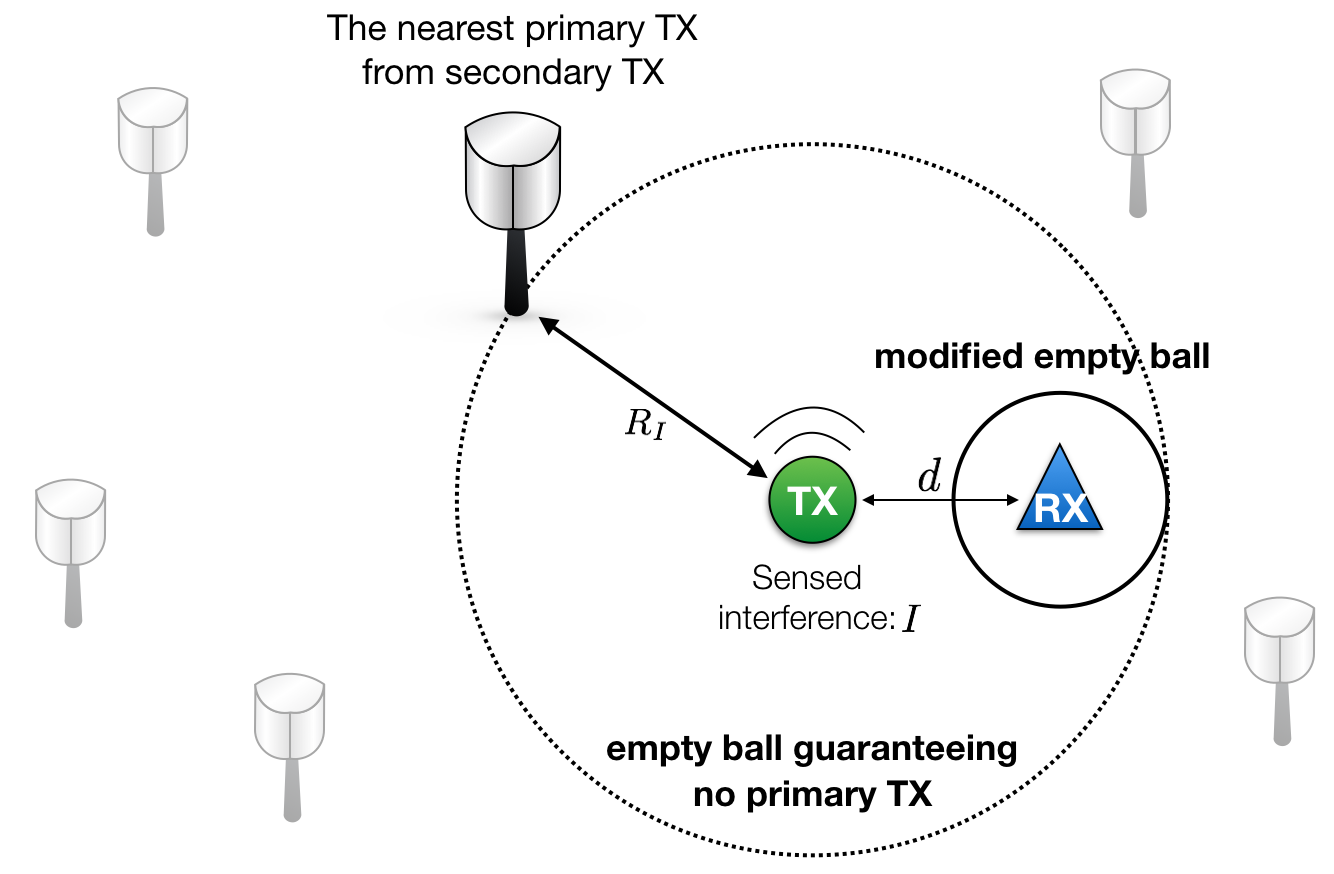}} 
\vskip -10pt
\caption{The sensed interference $I$ infers the distance $R_I$ as well as an empty ball condition.\label{derivation}}
\vskip -15pt 
\end{figure}

\subsection{Sense-and-Predict}

Time is slotted and each slot comprises sensing and transmission periods, which is synchronized among secondary TXs. 
At the sensing period, every secondary TX measures the aggregate interference from primary TXs via the energy detection method \cite{Yucek09}. 
It is assumed that the sensing period is large enough to guarantee the accurate measurement. Let $I^{(\ell)}$ denote the level of primary interference at the location of secondary TX $\ell$.  
Given the measured value of $I^{(\ell)}$,  
it is possible to predict the level of primary interference at its paired RX 
that is spatially separated with a distance of $d$. 
Specifically, secondary TX $\ell$ calculates the conditional coverage probability that given the measured interference $I^{(\ell)}$, its received $\SIR_2$ 
is larger than a predetermined access threshold $\theta$ 
while the other secondary TXs are assumed to be silent 
($\phi_2^{(i)}=0,  \forall i\neq \ell$ in \eqref{Eq:SIR_Definition}). 
The conditional probability is utilized as the access probability of secondary TX~$\ell$, denoted by $\phi_2(I^{(\ell)},\theta)$. 

At the transmission period, secondary TX $\ell$ transmits its packet 
with probability $\phi_2(I^{(\ell)},\theta)$ like ALOHA systems \cite{aloha}. 
The transmission is said to be successful when the received $\SIR_2$ is larger than a given decoding target $\beta$.
Note that other secondary TXs utilize different access probabilities $\phi_2(I^{(i)},\theta)$, followed by their measured interference level.

 \begin{figure*}[b]
\hrulefill
{\small
\begin{align}\nonumber
&\mathcal{P}_s(R_I,\ \theta) =  \frac{1}{2\pi} \int_{0}^{2\pi}\frac{P_2}{P_2+{P_1 \theta} d^\alpha \left(R_I^2-2 d R_I \cos(t) + d^2\right)^{-\frac{\alpha}{2}}} dt \\ &\times
 \exp\left(-\lambda_1\[\int_{R_I+d}^{\infty}\frac{2\pi{P_1 \theta} d^\alpha y^{-\alpha+1}}{P_2+{P_1 \theta} d^\alpha y^{-\alpha}}dy+\int_{|R_I-d|}^{R_I+d} \frac{2\text{acos}\left(\frac{R_I^2-d^2-y^2}{2 d y}\right){P_1 \theta} d^\alpha y^{-\alpha+1}}{P_2+{P_1 \theta} d^{\alpha} y^{-\alpha}} dy + \int_{\max(0,R_I-d)}^{|R_I-d|}\frac{2\pi{P_1 \theta} d^\alpha y^{-\alpha+1}}{P_2+{P_1 \theta} d^\alpha y^{-\alpha}}dy\]\right) \label{Eq:CondSIR1} 
 \end{align}}
\end{figure*}

\begin{figure*}[b]
\hrulefill
{\small
{\begin{align}\mathcal{P}_s(R_I,\theta) \geq  \begin{cases} 
\frac{1}{2\pi} \int_{0}^{2\pi}\frac{P_2}{P_2+P_1\theta d^\alpha \left(R_I^2-2 d R_I \cos(t) + d^2\right)^{-\frac{\alpha}{2}}} dt \times \exp\(-\pi\lambda_1\( \frac{P_1}{P_2}\)^\frac{2}{\alpha} d^2 \rho_0(\theta)\) 
& {\textrm{if $  d > R_I$,}} \\  \frac{1}{2\pi} \int_{0}^{2\pi}\frac{P_2}{P_2+P_1\theta d^\alpha \left(R_I^2-2 d R_I \cos(t) + d^2\right)^{-\frac{\alpha}{2}}} dt \times
 \exp\left(-\pi \lambda_1 {(R_I-d)}^2 \left[\rho \left(1+\frac{d^\alpha P_1 \theta}{P_2 {(R_I-d)}^\alpha}\right)^{\frac{2}{\alpha}}-1\right] \right)
 &{\textrm{{if $d \leq R_I$,} }}  
\end{cases} \quad
 \label{Eq:SIR_lb}
\end{align}}}
where $\rho_0(x):=x^{\frac{2}{\alpha}}\int_0^\infty {\frac{du}{1+u^{\frac{\alpha}{2}}}}$ and $\rho=\frac{2\pi}{\alpha}\csc\(\frac{2\pi}{\alpha}\)$. 
\end{figure*}

\subsection{Problem Formulation}
Let us define area spectral efficiency (ASE) $\mathcal{A}$ as the sum of data rates of secondary RXs per unit bandwidth in a unit area \cite{Alouini99}.
We aim at maximizing the ASE while guaranteeing the primary network protection by solving the following optimization:
\begin{align}\nn  
&\underset{\theta,\beta}{\text{max}}\ \mathcal{A}:=  \lambda_2 \mathsf{E}_I [{\phi}_2 (I,\theta)\mathsf{P}(\SIR_2>\beta |I)    ]  \ln(1+\beta),\\
&\ \ \text{s.t. }\ \mathsf{P}[\SIR_1 \leq \gamma] \leq \tau  \label{Eq:ASE_cond} \tag{P1},
\end{align} \normalsize
where the term $\mathsf{P}(\SIR_2>\beta|I)$ means the transmission success probability with a decoding target $\SIR$ $\beta$ conditioned on  the measured interference level of $I$ at the secondary TX. 
The constraint represents the protection condition of the primary network where $\SIR$ outage probability of a typical primary user does not exceed a given constraint $\tau$.
It is worth mentioning that both of the access threshold $\theta$ and decoding target $\beta$ are represented as $\SIR$s but their corresponding usages are distinct.
First of all, the access threshold $\theta$ determines whether to access the spectrum or not according to the interference level from primary TXs.
On the other hand, the decoding target $\beta$ determines if the received signal can be decoded through the interference from primary and secondary TXs.

\vskip -5pt
\section{Opportunistic Spectrum Access  on Sense-and-Predict}

In the SaP framework, a secondary TX tries to access the spectrum with probability $\phi_2(I,\theta)$,
which depends on the level of interference correlation between TX and RX.   
The specified derivations and its mathematical forms are described firstly, and then its verifications through both  simulation and USRP testbed are presented.

\subsection{Analysis of Access Probability}

Actually, the direct derivation of the access probability $\phi_2(I,\theta)$ is too complicated to make a tractable expression 
since the secondary TX and RX share common interferers causing the angular correlation.
Such correlation violates the property of isotropy in point process, blocking the use of standard techniques of PPP such as Campbell's theorem \cite{Haenggi13}.
To address this issue, we make the following assumption.
\begin{assumption}\label{EmptyBallAssumption}
\emph{ (Empty Ball)
Assume that there is an empty ball with radius of $R_I$ on which the nearest TX is located 
as shown in Fig.~\ref{derivation}. 
In other words, no primary TX exists inside the empty ball. 
Furthermore, the other primary TXs located outside the empty ball are assumed to follow 
PPP. 
}
\end{assumption}
Assumption \ref{EmptyBallAssumption} enables to decompose the access probability $\phi_2(I,\theta)$ into the product of two terms:
The first term is affected by the interference from the nearest primary TX on the empty ball, of which the angular correlation is expressed using a single trigonometrical function. 
The other is affected by the aggregate interference from the primary TXs outside the empty ball, of which the angular correlations are disregarded because of the isotropic property in PPP.  
The main result is shown in the following proposition.

 \begin{proposition}\emph{(Access Probability) 
Based on Assumption \ref{EmptyBallAssumption}, the access probability is given as $\phi_2(I,\theta) = \mathcal{P}_s (R_I,\theta),$ where the probability $\mathcal{P}_s (R_I,\theta)$ is given as \eqref{Eq:CondSIR1} at the bottom of the next page and the distance $R_I$ satisfies the following equation:
{{\begin{align}\label{Eq:EmptyBallRadius}
\frac{I}{P_1} {{R_I}}^\alpha-\frac{2\pi\lambda_1}{\alpha-2}{{R_I}}^2-1=0.
\end{align}}\normalsize}
Furthermore, when the path-loss exponent $\alpha$ is given as $4$, $R_I$ is reduced to {$(2I)^{-\frac{1}{2}}\({{\pi\lambda_1 P_1 + \[{\(\pi\lambda_1 P_1\)^2+4P_1 I}\]^{\frac{1}{2}}}}\)^{\frac{1}{2}}$}. 
\vskip 0pt \noindent
{\bf  Proof.} 
See Appendix.
 \hfill $\blacksquare$}
\end{proposition}

Note that the access threshold $\theta$ decides to access the spectrum aggressively or conservatively.
As mentioned above, the spectrum access opportunity varies with the network parameters such as primary TX density, even the interference $I$ is identical.
It is thus of great importance to optimize such access threshold, which will be provided in the following section.


It is notable that the radius of the empty ball $R_I$ in \eqref{Eq:EmptyBallRadius}, 
playing an important role to predict the level of the interference at the secondary RX,   
depends on different network parameters as specified in the following remark.

\begin{remark} \emph{ (Radius of the Empty Ball) 
The distance $R_I$ varies with different network parameters:
\begin{enumerate}[$\hspace{10pt}1.$]
\item{$R_I$ decreases with the measured interference $I$.}
\item{For high density $\lambda_1$ and/or power $P_1$, the aggregate interference from primary TXs outside of the empty ball is high. This increases the distance $R_I$ to reduce the interference power from the nearest primary TX and thereby maintain the interference $I$, and vice versa for the opposite situation. }
\item{For a high path-loss exponent $\alpha$, the aggregate interference from TXs outside of the empty ball decreases. 
The distance $R_I$ thus should be shorter to capture the effect that the nearest primary TX becomes more dominant.  }
\end{enumerate}
} 
\end{remark}

According to Proposition 1, furthermore, the access probability has remarkable characteristics about the measured interference $I$ in asymptotic cases as in the following remark.

\begin{remark} \emph{(Asymptotic Access Probability) In the following two asymptotic cases, the access probability $\phi_2(I,\theta)$ has distinct characteristics:
\begin{enumerate}[$\hspace{10pt}1.$]
\item{For $R_I\gg d$, i.e. for small $I$, the access probability becomes
\small\begin{align}
\frac{P_2 \exp\[{-\pi\lambda_1\(\frac{P_1\theta d^\alpha}{P_2}\)^{\frac{2}{\alpha}}\int_{\(\frac{P_2 {R_I}^2}{P_1 \theta d^\alpha}\)^{\frac{2}{\alpha}}}^\infty {\frac{du}{1+u^{\frac{\alpha}{2}}}}}\]}{P_2+P_1\theta\(\frac{d}{R_I}\)^2},
\end{align} \normalsize
which is equivalent to the $\SIR$ coverage at the secondary TX whose measured interference is $I$. 
As $I$ decreases to $0$, this probability converges to $1$, $\lim_{I\rightarrow 0} \phi_2(I,\theta)=1$.}
\item{For $R_I\ll d$, i.e. for large $I$, the access probability converges to 
\small\begin{align}
\lim_{I\rightarrow \infty}\phi_2(I,\theta)\!\!=\!\!\frac{P_2 \exp\({-\pi\lambda_1\(\frac{P_1\theta d^\alpha}{P_2}\)^{\frac{2}{\alpha}}\int_0^\infty {\frac{du}{1+u^{\frac{\alpha}{2}}}}}\)}{P_2+P_1\theta} ,
\end{align} \normalsize
which is totally independent of $I$.}
\end{enumerate}}
\end{remark}

Such arguments stem from that the dependency of the aggregate interference between the secondary TX and RX is weakened as $I$ increases.
The reason behind is that the empty ball radius decreases with $I$ (see Remark 1), yielding that the nearest primary TX of the secondary TX and that of secondary RX cannot be identical any more with high probability.
In this case, the measured interference $I$ that is dominantly dependent on the nearest interferer of the secondary TX has a low impact on the interference level at RX.
This remark is visualized in Fig.~\ref{fig:CondSIR_ri}, which will be explained in the following subsection.

To provide more applicable SaP design guideline, we consider a less tight but more tractable bound of the access probability by exploiting the following assumption.

\begin{assumption}\label{ModifiedEmptyBallAssumption}
\emph{ (Modified Empty Ball)
Without the change of the nearest primary TX location,  
no primary TX exists inside the modified empty ball: 1) located on the secondary RX and 2), of which the radius is $\max(0, R_I-d)$ as shown in Fig.~\ref{derivation}.
}
\end{assumption}

\begin{figure}     
\centering
   \subfigure[Without SaP]{\centering
     \includegraphics[width=4.25cm]{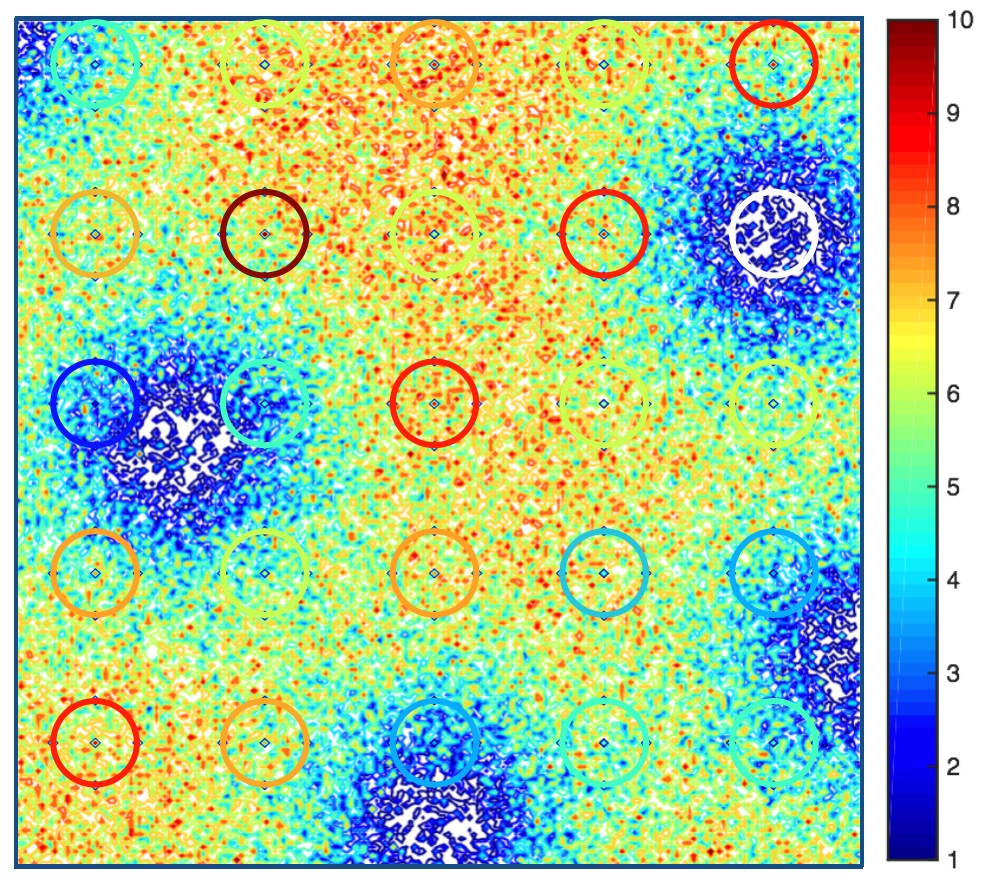} } 
   \subfigure[With SaP]{
     \includegraphics[width=4.25cm]{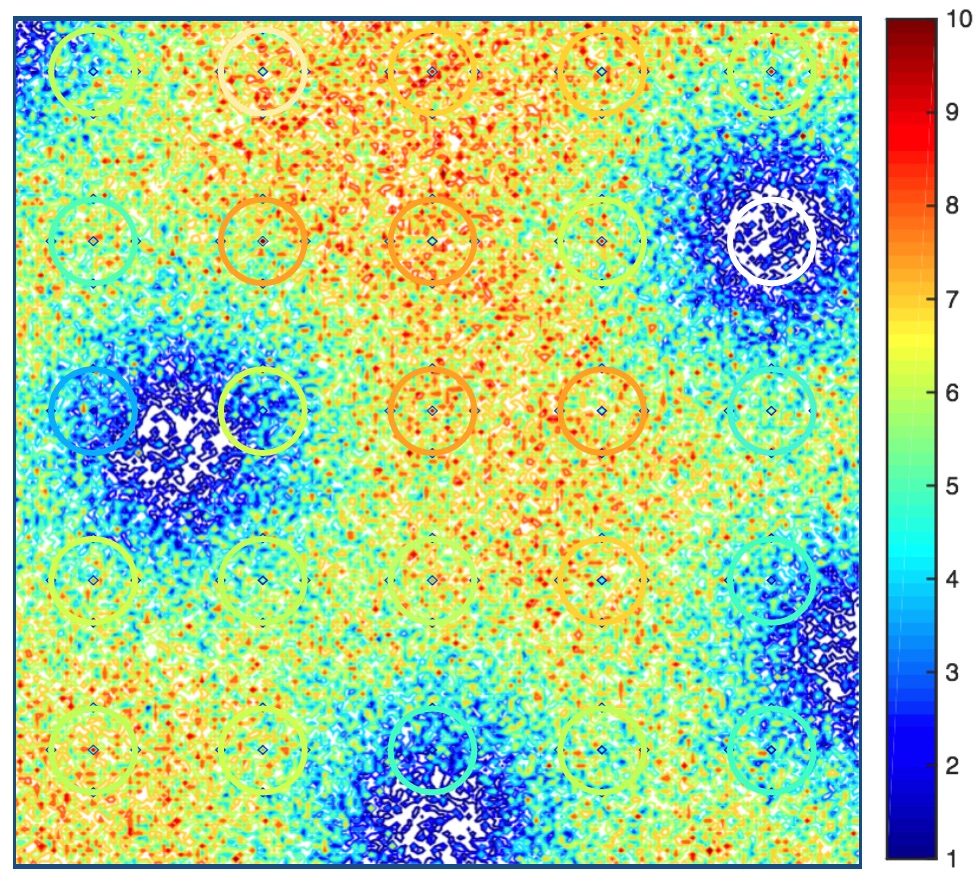}} 
     \vskip -5pt
\caption{The $\SIR$ prediction where $\lambda_1= 10\text{ TXs}/\text{km}^2$ and $d$ = 2 m. When the center of each circles denote a secondary TX location and its circumference the possible pair RX locations, the colors of them represent the sensed $\SIR$ at TX and predicted $\SIR$ at RX, respectively.\label{Fig:SIRcompare}}
\vskip -10pt
\end{figure}

According to this assumption, the first integration of the access probability \eqref{Eq:CondSIR1} remains the same since the nearest primary TX from the secondary TX is fixed.
On the other hand, the second exponential term has to be changed according to the modified empty ball located on the secondary RX.
With such modified empty ball condition, no primary TX is guaranteed within the distance of $\max(0,R_I-d)$ from the secondary RX.

%

 \noindent\begin{proposition}\emph{(Access Probability Lower Bound) The access probability is lower bounded as \eqref{Eq:SIR_lb} at the bottom of the previous page. Note that the equality holds when $d=0$.
\vskip 0pt
\noindent
{\bf  Proof.} 
See Appendix.
 \hfill $\blacksquare$}
\end{proposition}

The advantage of this simplified access probability lower bound is in reduction of computational complexity. 
Specifically, the access probability can be decomposed into the terms of $R_I$ and others, and each secondary TX thereby only needs to update the value of $R_I$ that is varying each time slot. Remind that without this bound, each secondary TX should calculate the whole integral expression of \eqref{Eq:CondSIR1} in every slot.

\subsection{Verification of Access Probability}

In order to verify the accuracy of Proposition 1, we conduct the following experiment as shown in Fig. \ref{Fig:SIRcompare}.
Consider that the density of primary TXs $\lambda_1$ is $10 \text{ TXs}/\text{km}^2$ and their transmit power is $43$ dBm.
We pick $25$ points as candidate locations of the secondary TXs and their colors represent measured interference levels, more precisely $\SIR$s (See the centers of circles in Fig.~\ref{Fig:SIRcompare}).
To represent the interference level as an $\SIR$, we assume the received signal power is fixed to be $10$ dBm and calculate the corresponding $\SIR$.
Each point has a circle with a radius of $2$ meters, where the circumference represents every possible secondary RX location 
when $d=2$ (in meters) and the color means the predicted $\SIR$ depending on the MAC protocol.
Without the SaP, the color of center and circumference are the same because 
a secondary TX utilizes its measured value directly (see Fig.~\ref{Fig:SIRcompare} (a)). 
When utilizing the SaP, on the other hand, the color of circumference is determined by the average value of 
$\SIR$ derived by Proposition 1, and the center and circumference may have different colors as shown in Fig.~\ref{Fig:SIRcompare}~(b). 
It is observed that with SaP, most of circles become invisible whereas circles without the SaP are easily discovered, verifying that the proposed SaP can predict interference at RX properly, and thereby decreasing errors such as missed detection and false alarm.
In addition, some circles remain unsolved (see a white circle in Fig.~\ref{Fig:SIRcompare}~(a)~and~(b)).
Such a white circle implies that the SaP hardly predict the interference at RX if the measured interference is too high. 
\begin{figure}
\centering 
{\includegraphics[width=7.5cm]{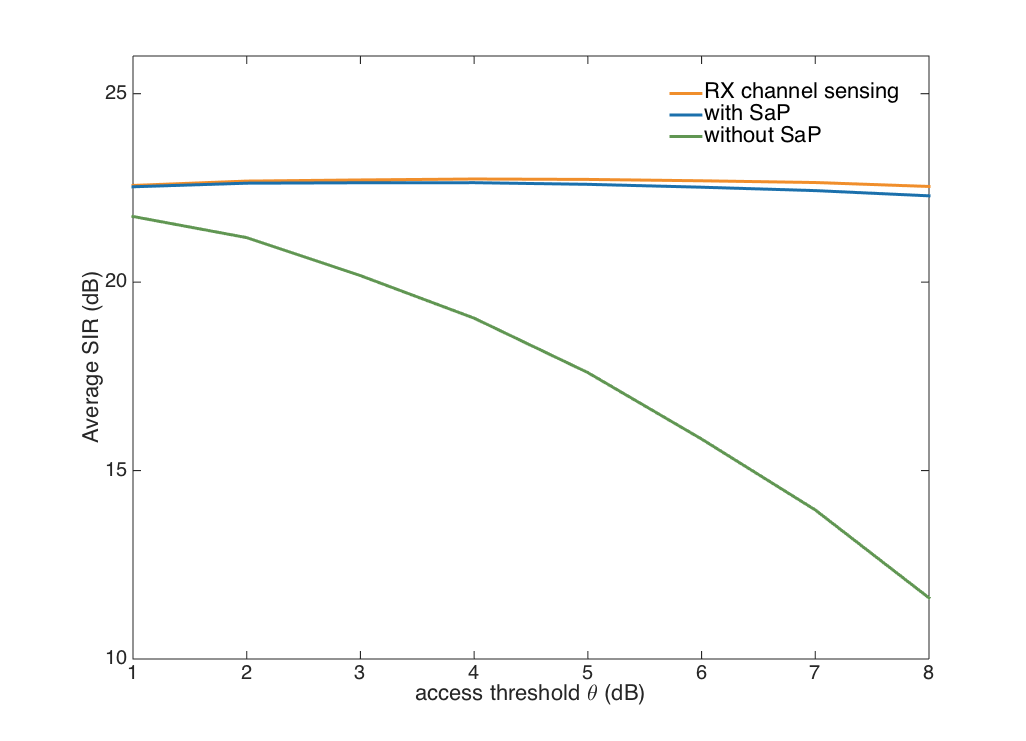}} 
\vskip -10pt
\caption{Comparison of average $\SIR$ when considering the RX channel sensing as an upper case ($\lambda_1=5 \times {10}^2 \text{ TXs}/\text{km}^2$, $P_1=43$ dBm, $P_2=23$ dBm).\label{fig:avgSIR} }
\vskip -15pt 
\end{figure}

Fig.~\ref{fig:avgSIR} represents the average value of the $\SIR$ according to the access threshold $\theta$,
showing that the gap between the average $\SIR$ with and without SaP increases with the access threshold.
This is mainly because high $\theta$ discourages secondary TX to access the spectrum.
The limited transmission attempts makes it difficult to recover the average $\SIR$ once the secondary RX receives a poor $\SIR$ due to the interference prediction failure.
Note that the channel sensing carried out at the secondary RX provides 
the upper bound performance (see the yellow line) in the sense that there is no prediction error. 
It is verified that SaP can achieve sufficient benefit in spectrum access without a direct channel sensing at the secondary RX.

\begin{figure}
\centering 
{\includegraphics[width=6.5cm]{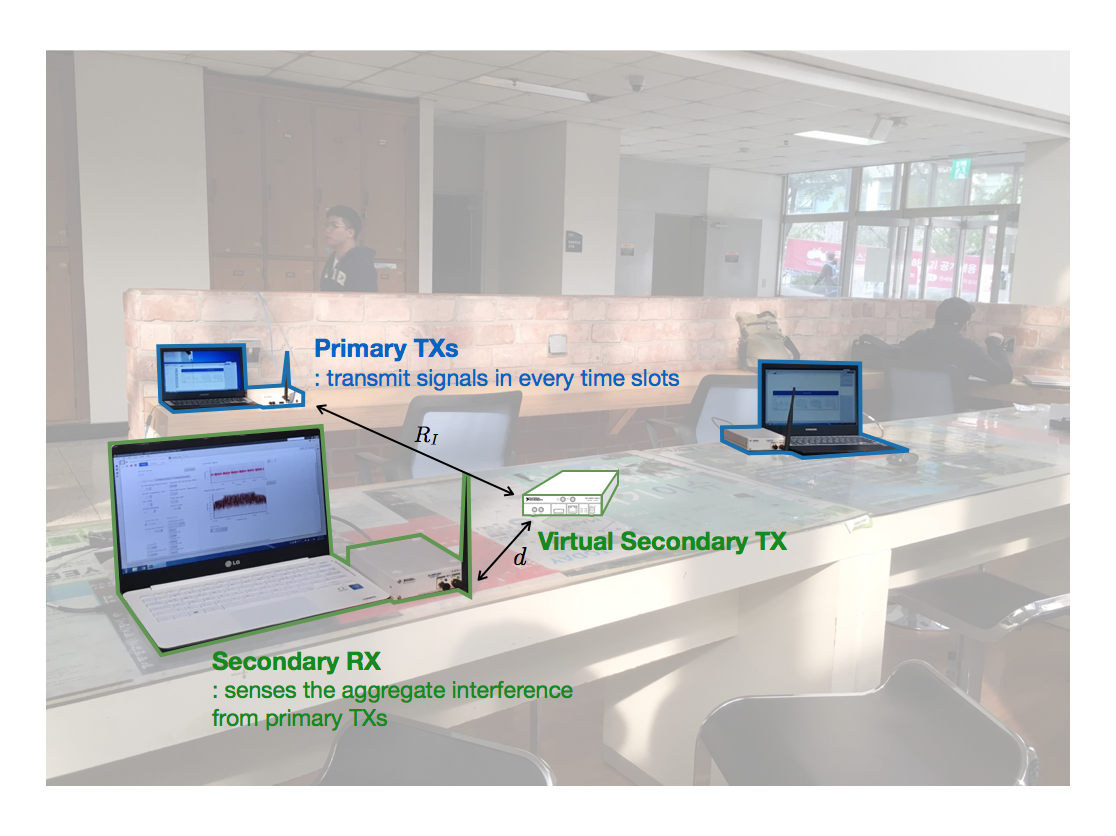}} 
\vskip -10pt
\caption{The USRP testbed setup for CR networks. \label{usrp}}
\vskip -15pt 
\end{figure}

To validate the analytic results of access probability $\phi_2(I,\theta)$ in Propositions 1 and 2, we compare them with the actual measurement results based on an experimental testbed using three laptops and three USRPs (see Fig.~\ref{usrp}).
The primary network comprises two USRPs as two primary TXs
where each of them transmits a signal with $11.3$ dBm at a center frequency of $1.3$ GHz in every time slot with the bandwidth of $200$ kHz.  
The secondary network comprises one USRP as a secondary RX.
Due to the limited number of USRP devices, we assume that there is a virtual secondary TX transmitting signal with a power $5$ dBm at the same carrier frequency with the primary TXs.
The laptops play a role to configure the paired USRP devices.  
The distance $R_I$ between this secondary TX and its nearest primary TX is $3.6$ m. 
The secondary RX measures the aggregate interference from primary TXs with the sensing bandwidth of $600$ kHz, and passes the signal to the lowpass filter with the bandwidth of $400$ kHz.  
Its receiver gain is $61$ dB.
With the measured interference level, 
the secondary RX calculates its $\SIR$ assuming that the signal from the virtual secondary TX only experiences the path-loss attenuation.

\begin{figure}
\centering 
{\includegraphics[width=7.5cm]{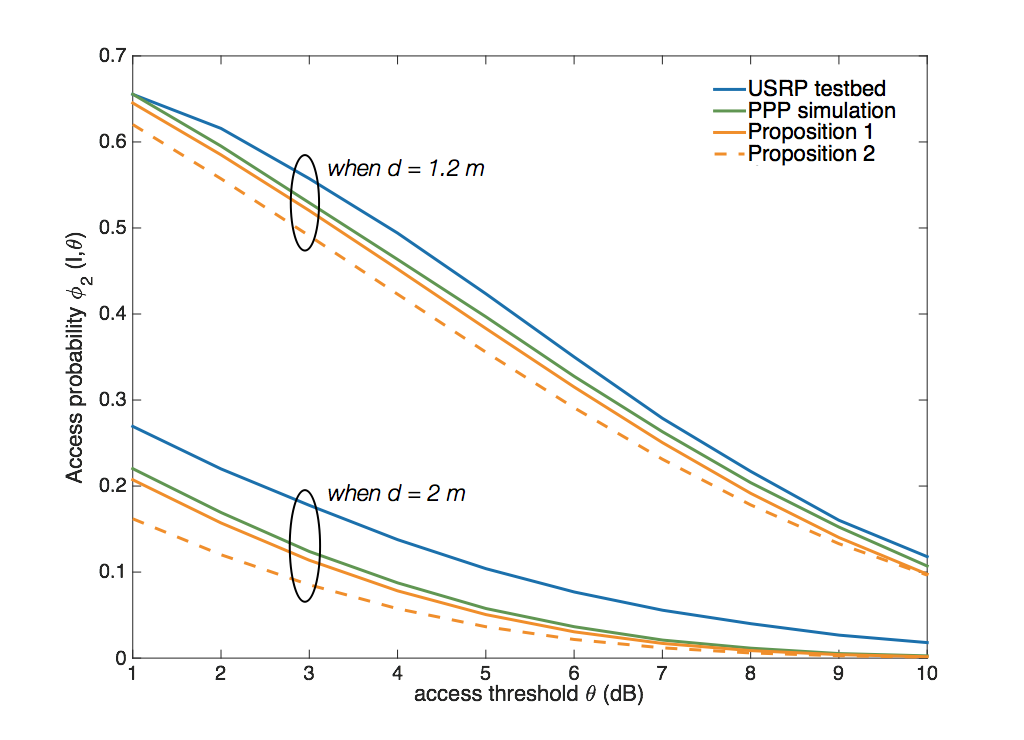}}
\vskip -10pt 
\caption{Access probability with SaP according to threshold $\theta$ ($R_I = 3.6 \text{ m}$, $\lambda_1= 7 \times 10^3 \text{ TXs}/\text{km}^2$, $P_1=11.3$ dBm, $P_2=5$ dBm, $\alpha=3$). \label{fig:CondSIR}} 
\vskip -20pt
\end{figure}

We also compare the analytic results with simulations. 
We consider a square of $1\times1$ $\text{km}^2$ where the primary TX density $\lambda_1$, transmit powers $P_1$ and $P_2$, and the distance $R_I$ are set to be the same as those of the USRP testbed experiment.
The path-loss exponent is $3$ by considering that the testbed is processed indoors.
Fig.~\ref{fig:CondSIR} shows that the analytic access probabilities from Proposition 1 and 2 are quite consistent with that from PPP simulation.
There is a slight difference between analytic and USRP experiment results but they follow similar tendencies.
The reason why the access probabilities from USRP testbed are higher than those from analysis is because two primary TXs cannot sufficiently represent the real primary network comprising almost infinite number of TXs.
This makes the secondary RX measures the aggregate interference lower than those in analysis and PPP simulation, yielding the increment in $\SIR$ coverage. 
This gap can be mitigated by increasing the USRP number and/or the experiment area.
 
In addition, the result indicates that the gap between access probability and its lower bound is bigger when the pair distance is $2$ m compared to the case of $1.2$ m.
This gap is mainly caused by the empty ball modification as in Fig.~\ref{derivation}.
Since the area difference between empty balls in Assumptions 1 and 2 increases with the pair distance, we can get a tight bound of the access probability when the pair distance is short.

\begin{figure}
\centering 
{\includegraphics[width=7.5cm]{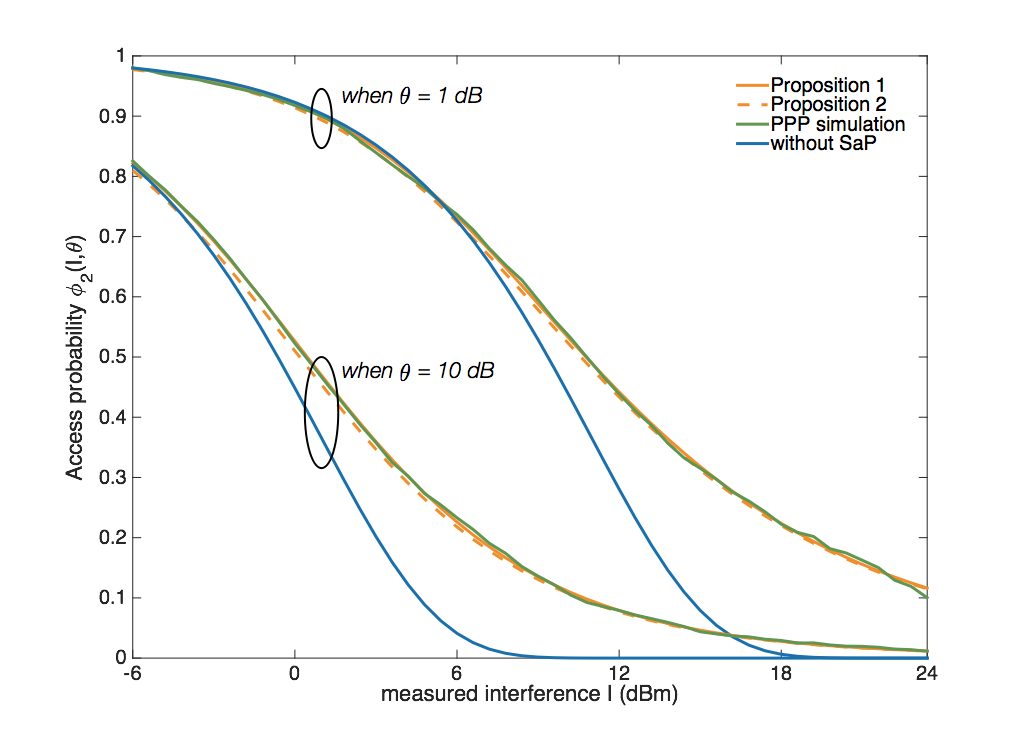}}
\vskip -10pt 
\caption{Access probability with SaP according to the measured interference $I$ ($d = 2$ m, $\lambda_1= 5 \times 10^2 \text{ TXs}/\text{km}^2$, $P_1=43$ dBm, $P_2=23$ dBm, $\alpha=4$). \label{fig:CondSIR_ri}}
\vskip -20pt 
\end{figure}

Fig.~\ref{fig:CondSIR_ri} visualizes the access probability according to the measured interference $I$.
As $I$ increases, it shows the access probability decreases, validating the analytic results are in quite good agreements with the simulated results.
Besides, the access probability gap between with and without SaP increases with $I$ due to the weakened interference correlation between secondary TX-RX (see Remark 2). 
This implies that the SaP can reduce the number of false alarms where the secondary TX decides not to access the spectrum due to the highly measured interference level, even if there is an access opportunity.

\vskip -5pt
\section{Area Spectral Efficiency Maximization}

\begin{figure*}     
\centering
   \subfigure[ASE (analysis)]{\centering
     \includegraphics[width=7.5cm]{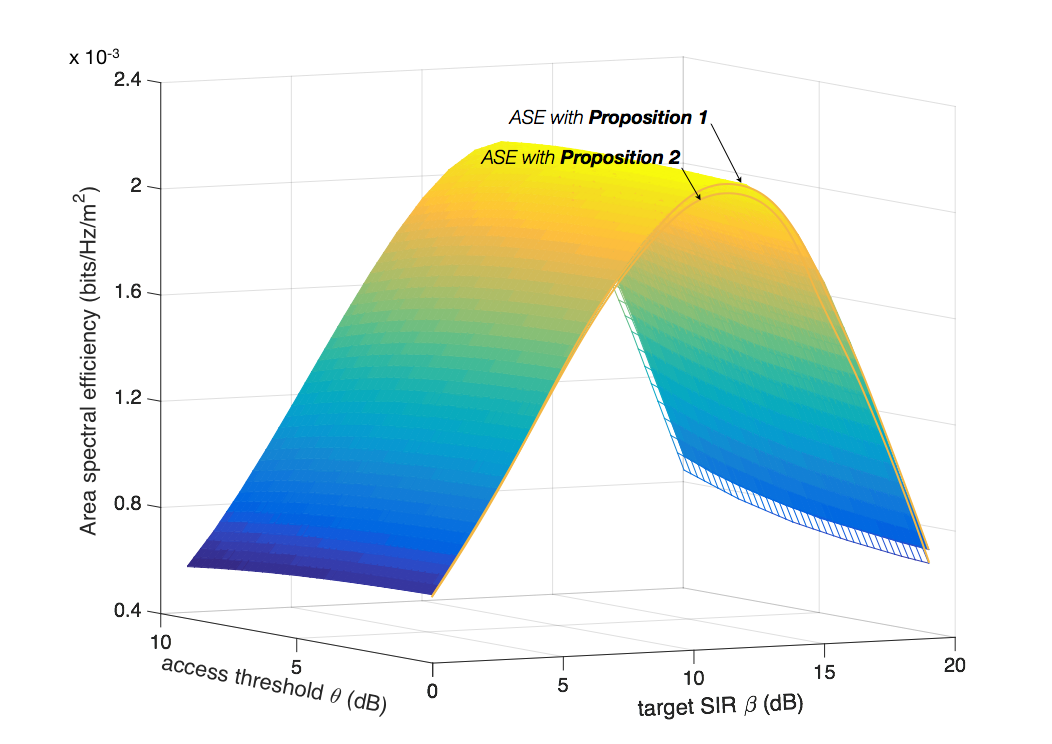} } 
   \subfigure[ASE (simulation+analysis)]{
     \includegraphics[width=7.5cm]{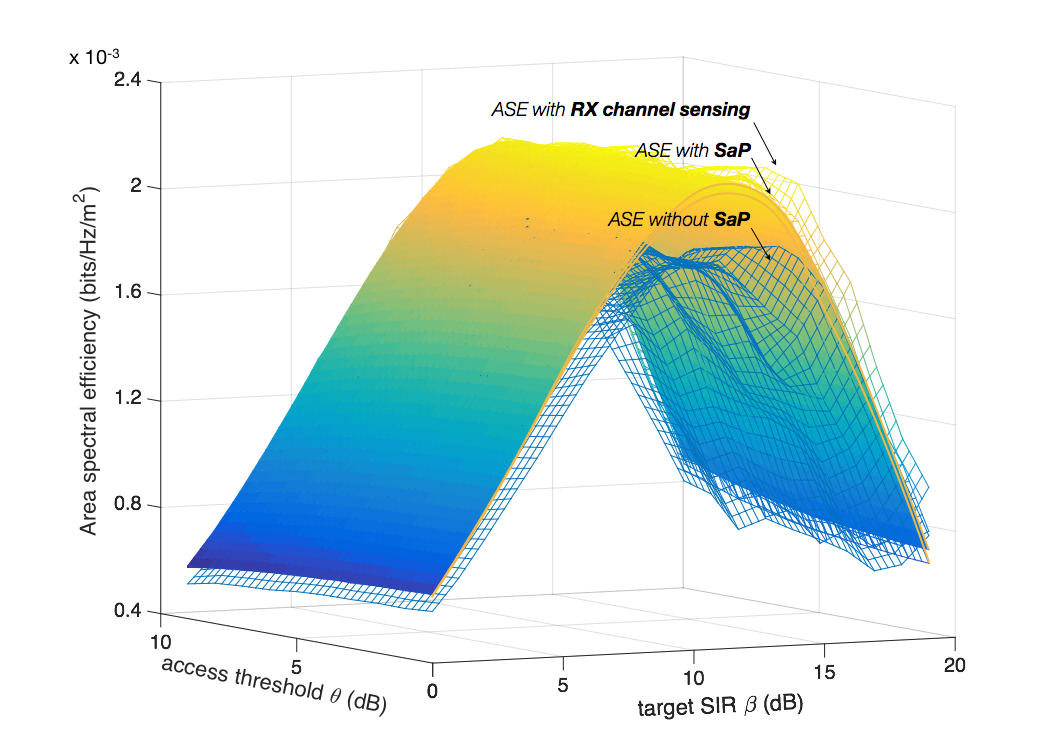}}
     \vskip -5pt
\caption{ASE analysis and simulation ($\lambda_1=500 \text{ TXs}/\text{km}^2$, $\lambda_2=200 \text{ TXs}/\text{km}^2$, $P_1=43$ dBm, $P_2=23$ dBm, $\alpha=4$).\label{Fig:ASE}}
\vskip -20pt
\end{figure*}

\begin{figure*}[b]
\hrulefill
{\small
\begin{align}
\mathsf{D}(\beta) = \int_0^\infty \mathcal{P}_s(r,\overline{\theta}) f_r(r) \[\mathcal{P}_s(r,\beta)\(-\frac{2 c_0 \beta^{\frac{2-\alpha}{\alpha}} }{\alpha}+\frac{1}{(1+\beta)\ln\(1+\beta\)}\)+\mathcal{P}'_s(r,\beta)\] dr
 \label{disc_D}
\end{align}}
where $c_0:= \pi\lambda_2 \hat{\phi}_2 d^2 \int_0^\infty \frac{du}{1+u^{\frac{\alpha}{2}}},$ and $\mathcal{P}'_s(r,\beta)$ is a differential function of $\mathcal{P}_s(r,\beta)$ lower bound \eqref{Eq:SIR_lb} with respect to $\beta$.
\end{figure*}

This section deals with ASE maximization of Problem \ref{Eq:ASE_cond} using Propositions~1~and~2. 
It is worth clarifying that even though both $\theta$ and $\beta$ are the threshold of $\SIR_2$, 
the optimal thresholds $\theta^*$ and $\beta^*$ should be derived separately   
because their associated probabilities are different. 
The access probability $\phi_2(I, \theta)$, associated with $\theta$, 
 considers interference from primary TXs only, and it is already expressed in form of $\mathcal{P}_s(R_I, \theta)$ in \eqref{derivation}. 
On the other hand, the coverage probability  $\mathsf{P}(\SIR_2>\beta|I)$, associated with $\beta$,
considers the aggregate interference from primary and secondary TXs but no mathematical form exists so far. 
It is thus required to express $\mathsf{P}(\SIR_2>\beta|I)$ in form of $\mathcal{P}_s(R_I, \theta)$ as a preliminary step.

The transmission success probability depends on the number of concurrent secondary TXs accessing the spectrum in a given time slot. 
For mathematical tractability, we approximate the concurrent secondary TXs are independently thinned with the average access probability.
The derivation of average access probability is specified in the following Lemma 1, enabling the expression of the $\mathsf{P}(\SIR_2>\beta|I)$ as given in Lemma 2.


 \noindent\begin{lemma}\emph{ (Average Access Probability) Given the access threshold $\theta$, 
the average access probability $\hat{\phi}_2(\theta)=\mathsf{E}_I [{\phi}_2 (I,\theta)]$ is
$\hat{\phi}_2(\theta)=\int_0^\infty {\mathcal{P}_s (r,\theta)} f_r(r) dr$, where $f_r(r)$ is the probability density function of the distance to the nearest primary TX, i.e. $f_r(r)=2\pi\lambda_1 r e^{-\pi\lambda_1 (r)^2}$.
\vskip 0pt \noindent
{\bf  Proof.} See Appendix. 
\hfill $\blacksquare$}
\end{lemma}

\noindent\begin{lemma}\emph{ \normalsize(Transmission Success Probability) \normalsize
Given the access threshold $\theta$ and the radius of the empty ball $R_I$, the decoding target $\beta$ and the measured interference $I$, the probability $\mathsf{P}(\SIR_2>\beta|I)$ is given as
{\begin{align}\nn
\mathsf{P}(\SIR_2>\!\!\beta|I)=\mathcal{P}_s(R_I,\beta) \exp\!\!\l({-\pi\lambda_2{\hat{\phi}_2 (\theta)}^{\frac{2}{\alpha}}d^2 \rho_0(\beta)}\r) .
\end{align}}\normalsize 
\vskip 0pt \noindent
{\bf  Proof.} See Appendix.
\hfill $\blacksquare$}
\end{lemma}

In addition, we consider a primary network protection condition which determines the minimum access threshold $\overline{\theta}$ to assure that the outage probability of the primary network does not exceed the threshold $\tau$.

 \noindent\begin{lemma}\emph{ \normalsize(Minimum Access Threshold) The optimal access threshold $\theta^*$ should be no less then $\overline{\theta}$ that satisfies the following equation: 
{\begin{align}\nn
{\lambda_2 \hat{\phi}_2 (\theta) {P_1}^{\frac{2}{\alpha}} \rho_0(\overline{\theta}) (1-\tau) }\!=\!{\lambda_1 {P_2}^{\frac{2}{\alpha}} \(\tau\!+\! \rho(\overline{\theta})\tau-\rho(\overline{\theta})]\)}.
\end{align}}\normalsize
\vskip 0pt \noindent
{\bf  Proof.} See Appendix. 
\hfill $\blacksquare$}
\end{lemma}

Note that Lemma 2 enables to derive the expression of ASE and Lemma 3 provides the feasible region of the access threshold $\theta$, leading to the following problem formulation that is equivalent to Problem \ref{Eq:ASE_cond}.

\noindent\begin{proposition}\emph{(Area Spectral Efficiency Maximization) Problem \ref{Eq:ASE_cond} 
is rewritten as
\begin{align}
\{\theta^*, \beta^*\} = \underset{\theta>\overline{\theta},\beta}{\mathrm{argmax}}  
\int_0^\infty & \large(\mathcal{P}_s(r,\theta) \mathcal{P}_s(r,\beta) e^{-\pi\lambda_2 {\hat{\phi}_2 (\theta)}^{\frac{2}{\alpha}} d^2
 \rho_0(\beta)}\nonumber\\
& \ln(1+\beta) f_{r}(r) \large) dr.\label{ProblemReformulation}\tag{P2}\end{align}
\vskip 0pt \noindent
{\bf  Proof.} 
See Appendix.
\hfill $\blacksquare$}
\end{proposition}

It is difficult to derive the optimal access threshold $\theta^*$ and target $\SIR$ $\beta^*$ of Problem \ref{ProblemReformulation} in closed forms except the 
asymptotic case where the interference from secondary TXs is much smaller than that from primary TXs
as shown in the following proposition.   

\noindent\begin{proposition}\emph{When {$P_1\lambda_1 \gg P_2\lambda_2$}\normalsize, the optimal access threshold is $\theta^*=\overline{\theta}$ and the target $\SIR$ $\beta^*$ satisfies $\mathsf{D}(\beta)=0$ where $\mathsf{D}(\beta)$ is given as \eqref{disc_D} at the bottom.
\vskip 0pt \noindent
{\bf  Proof.} 
See Appendix.
\hfill $\blacksquare$}
\end{proposition}


Proposition 4 provides the SaP design guideline in a special scenario that is asymptotic but highly in accordance with the trend of CR networks, such as in the internet-of-things network based on LoRa \cite{lora} and body area network \cite{body} where $P_1 \gg P_2$, and the rural broadband service in TV white space band \cite{rural} where $\lambda_1 \gg \lambda_2$.


Besides such an asymptotic scenario, the mathematical form in \ref{ProblemReformulation} provides useful design guidelines steming from the intuition that there is an optimal density of concurrent secondary transmissions maximizing ASE by jointly optimizing the spatial reuse and network interference.
\begin{definition} \emph{(Optimal Concurrent Secondary Transmissions) The optimal density of secondary transmissions in a given slot is denoted by ${\lambda_2}^*$ which can be tuned with modification of the optimal access threshold $\theta$, i.e. ${\lambda_2}^* := \lambda_2 {\hat{\phi}_2}(\theta^*)$.
}
\end{definition}

Through Definition 1, we summarize the SaP design guideline according to different secondary network parameters, as in the following remarks.

\begin{remark} \emph{(Effect of the Secondary TX Density) The density $\lambda_2$ affects the third exponential term in \ref{ProblemReformulation}.
As $\lambda_2$ increases, 
\begin{enumerate}[$\hspace{10pt}1.$]
\item {The optimal access threshold $\theta^*$ should be increased to keep the optimal density ${\lambda_2}^*$;}
\item {The optimal target $\beta^*$ remains constant only if the optimal density ${\lambda_2}^*$ can be guaranteed (i.e., $\lambda_2 > {\lambda_2}^*$).
However, when the secondary TX density $\lambda_2$ is smaller than ${\lambda_2}^*$, there is no way to meet ${\lambda_2}^*$ by controlling $\theta$, decreasing $\beta^*$ to reflect the interference increment. }
\end{enumerate}
}
\end{remark} 

The impact of the secondary transmit power has a slight different characteristic as follows.

\begin{remark} \emph{(Effect of Secondary Transmit Power) Large power $P_2$ makes the first two probabilities in  \ref{ProblemReformulation} increase. 
Both of optimal access threshold $\theta^*$ and target $\beta^*$ thus should be increased, for maintaining the optimal density ${\lambda_2}^*$ and for reflecting the signal power increment, respectively.}
\end{remark} 

\begin{figure}
\centering 
{\includegraphics[width=7.5cm]{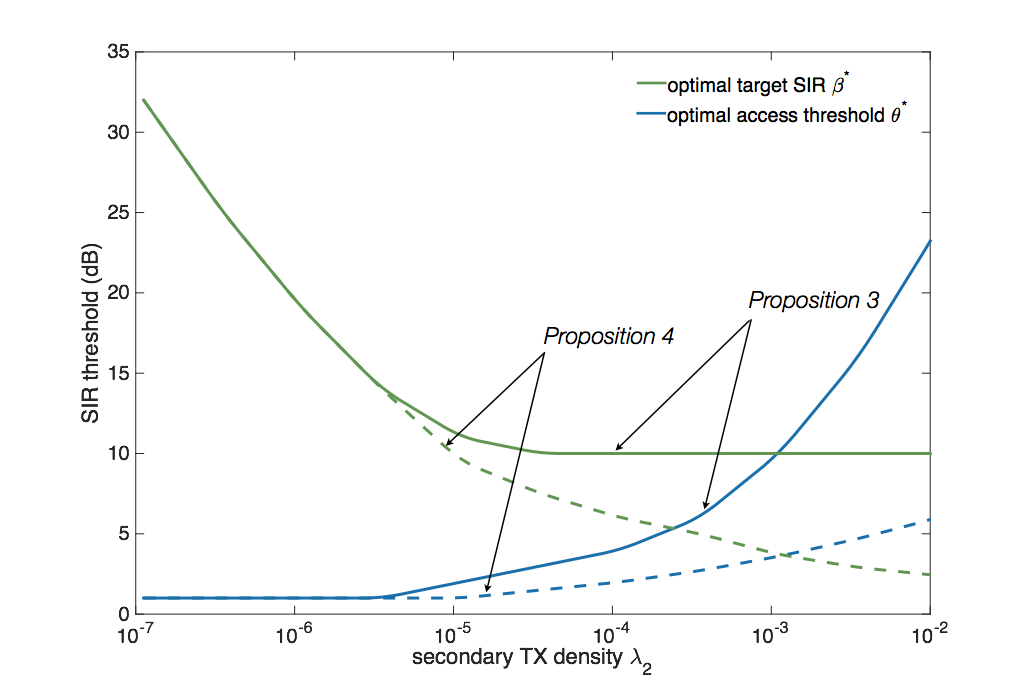}}
\vskip -10pt 
\caption{Optimal access threshold and target $\SIR$ according to the density $\lambda_2$ ($d = 2$ m, $\lambda_1= 5 \times 10^2 \text{ TXs}/\text{km}^2$, $P_1=43$ dBm, $P_2=23$ dBm, $\alpha=4$). \label{fig:ase_opt}}
\vskip -15pt 
\end{figure}

%
%

Fig.~\ref{Fig:ASE} shows that the ASE when utilizing the lower bound access probability \eqref{Eq:SIR_lb} in Proposition 2 matches well with ASE using \eqref{Eq:CondSIR1} in Proposition 1.
In addition, the optimal variables $\theta^*$ and $\beta^*$ of them are almost identical, implying that the secondary TX can utilize the access probability lower bound \eqref{Eq:SIR_lb} that effectively reduces the computation complexity.
This figure further shows that the analytic ASE is in a quite good agreement with the simulated one and thus validates the analysis.
When comparing to the simulated result of ASE without SaP, the proposed SaP can achieve much higher ASE, close to the ASE with RX channel sensing.

Fig.~\ref{fig:ase_opt} illustrates the optimal variables $\theta^*$ and $\beta^*$ according to the density $\lambda_2$.
As described in Remark 3, $\beta^*$ is saturated at some point of $\lambda_2$, while $\theta^*$ keeps increasing.
This figure also shows that the optimal variables derived from Proposition 4 fit well with those from linear searching (Proposition 3), although their gaps increase with $\lambda_2$. 
This implies that the network can utilize the Proposition 4 when $\lambda_2$ is low, for reducing the searching time in return for the accuracy decrement.

\section{Practical Considerations for the Design of SaP}
This section discusses some practical considerations in order to operate SaP in a real environment.

\begin{figure}
\centering 
{\includegraphics[width=7.5cm]{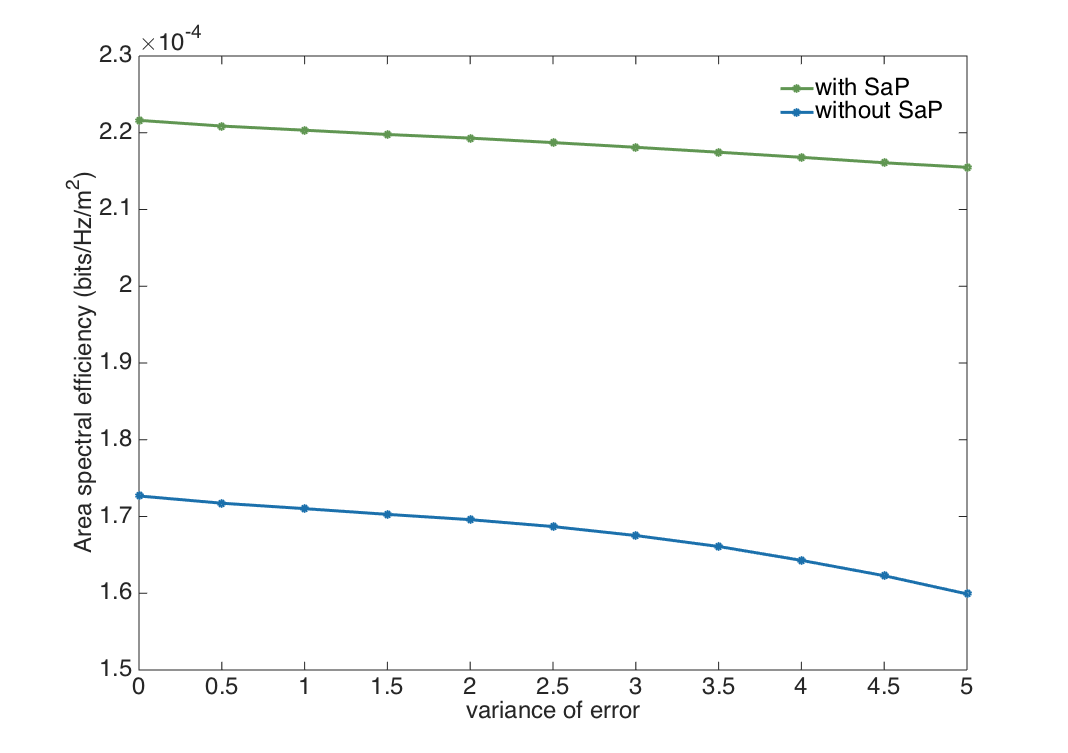}}
\vskip -10pt 
\caption{ASE according to the variance of interference measuring error ($\lambda_1= 5 \times 10^2 \text{ TXs}/\text{km}^2$, $\lambda_2= 2 \times 10^2 \text{ TXs}/\text{km}^2$, $P_1=43$ dBm, $P_2=23$ dBm). \label{fig:error}}
\vskip -15pt 
\end{figure}

\textbf{Interference Measurement.} In practice, the said interference measurement error can occur due to the thermal noise or some sensing circuit problems. 
It may result in miscalculating the spectrum access opportunity, 
and thereby decrease the CR network performance.
In order to verify the impact, we measure ASE by simulation when 
the measurement error is modeled as the normal distribution with zero mean, 
where the variance represents the level of the error.  
Fig.~\ref{fig:error} shows that SaP still achieves significant ASE than that without SaP.
In addition, ASE without SaP decreases more faster than that with SaP, 
implying that SaP is more tolerant to the interference measurement errors 
than the conventional TX sensing method using the level of measured interference directly for the access decision. 
On the other hand, the SaP  considers the primary network environment such as TX density or power, enabling to calibrate the incorrectly sensed interference in some degree.

\textbf{Path-loss Model.} Recall the fact that
all derivations in this paper entirely rely on the assumption that the path-loss model is given.
We used a simplified single-slope path-loss model where a large-scale signal attenuation depends on one path-loss exponent    
that is  affected by dynamic environmental changes, e.g. antenna selection, cell type or frequency band. 
It is indeed required to trace the exponent consistently 
via some algorithms by estimating the mean interference \cite{ple_haenggi} or an internode distance \cite{ple_fidan}.
In addition, it is worth considering extension to more realistic path-loss models as a multi-slope model or ray tracing  
that remains for future work.   

\textbf{Primary TX Density.} The primary TX density is another important factor we should estimate in order to not only figure out aggregate interference to a secondary RX but also guarantee the service quality of the primary user.    
There are two ways to acquire the primary TX density: i) periodical update by the network, e.g. DTV or LTE systems \cite{netmanias}; and ii) estimation based on statistical techniques such as a maximum likelihood method based on the received power strength \cite{onur}.

\vskip -10pt
\section{Conclusion}
This paper tackles the bottleneck of detecting the spectrum access opportunity in CR networks, due to the spatial difference between secondary TX and RX.
As its solution, we propose an SaP that predicts the interference level at the secondary RX based on the channel measurement at its paired TX.
The impact of the SaP is analyzed in terms of the secondary network ASE by using the stochastic geometry.
The result sheds light on the joint optimization of access threshold and target $\SIR$ in order to maximize the secondary network ASE while protecting the primary network.
Such optimal parameters are derived, providing the SaP design guideline.

The weakness of this study is that it does not consider the shadowing effect that intensifies the probability of miss detection and false alarm.
The SaP has a potential to further improve the accuracy of opportunity detection by estimating the geographic environment surrounding the secondary RX.
Further extension could also contemplate the impact of cooperative sensing among secondary TXs known to outperform the non-cooperative sensing approach in return for the increment in energy consumption \cite{sylee}.
The advantages of cooperative sensing and SaP are quite similar in that both of them enhance the accuracy of spectrum opportunity detection. 
But their operating behaviors are distinct: the accuracy of cooperative sensing decreases as the number of cooperation between secondary TXs decreases; whereas that of SaP decreases with the distance between the secondary TX-RX.
By jointly considering such relationships, we can provide the CR network solutions to maximize the spectrum utilization.

\renewcommand{\baselinestretch}{1}
\vskip -5pt \noindent 
\section*{Acknowledgment}
This work was supported by Institute for Information \& communications Technology Promotion grant funded by the Korea government (No.B0126-16-1017), and 'The Cross-Ministry Giga KOREA Project' grant from the Ministry of Science, ICT and Future Planning, Korea (No.GK 16S0400).


\vskip -5pt \noindent

\section*{Appendix}

\subsection{Proof of Proposition 1}

The desired result comes from the following two parts.

\subsubsection{Distance between a typical secondary TX and its nearest primary TX $T_1$}

Suppose that the distance between a typical secondary TX and its nearest primary TX is equal to $R_I$.
With this condition, the expected value of interference is represented as below with the aid of Campbell's theorem \cite{Haenggi13}.

\small\begin{align}
{P_1\mathsf{E}_h\[h {R_I}^{-\alpha}\]}+ \underbrace{P_1 \mathsf{E}_{\Phi_1, \{h^{(i)}\}}\[\sum_{i \in \Phi_1 \setminus T_1} h^{(i)} {x_1^{(i)}}^{-\alpha}\]} _{(a)}, \label{avg_intf}
\end{align}\normalsize
where $T_1$ denotes the nearest primary TX.
Let $I_r$ denote $\sum_{i \in \Phi_1 \setminus T_1} h^{(i)} {x_1^{(i)}}^{-\alpha}$ in $(a)$.
In order to derive $(a)$, we consider the Laplace transform of $I_r$, $\mathcal{L}(s) = \mathsf{E}\[e^{-s I_r}\]$.
{\small\begin{align} \nn 
\mathcal{L}(s) &= \mathsf{E}_{\Phi_1} \[\prod_{i\in \Phi_1 \setminus T_1} \mathsf{E}_{h^{(i)}} \(e^{-s h^{(i)} {x_1^{(i)}}^{-\alpha}}\)\] \\ \nn
&\overset{(b)}{=} \exp\(-2\pi\lambda_1\mathsf{E}_h\[\int_{R_I}^\infty \(1-\exp\(-s h r ^{-\alpha}\)\) r dr\]\) \\ \nn
&\overset{(c)}{=}\exp\(-\pi\lambda_1\mathsf{E}_h\[\int_0^{{R_I}^{-\alpha}}\frac{ t^{-\frac{2}{\alpha}} s h dt}{ e^{s h t}} - r^2 \(1-e^{-s h r^{-\alpha}}\)\]\)
\end{align}}\normalsize\noindent

\noindent where $(b)$ follows from the probability generating function of PPP \cite{Haenggi13} and $(c)$ from the partial integral.
By applying that $ \mathsf{E}\[I_r\] = -\frac{\partial}{\partial s}\mathcal{L}(s) |_{s=0}$, the function $(a)$ is represented as 
$
\frac{2 P_1 \pi\lambda_1 \mathsf{E}\[h\] {R_I}^{2-\alpha}}{\alpha-2}.
$
Now that the fading $h \sim \exp(1)$, finding $R_I$ that satisfies $I = P_1 {R_I}^{-\alpha} +\frac{2 P_1 \pi\lambda_1  {R_I}^{2-\alpha}}{\alpha-2}$ finalizes the proof.

\subsubsection{Access probability with an empty ball condition}

Let the value $x$ denote the distance between a typical secondary RX and $T_1$.
Then the access probability is represented as

\small\begin{align}
&\mathsf{P}\[\frac{P_2 h^{(0)} d^{-\alpha}}{P_1 \(h x^{-\alpha} + \sum_{i\in \Phi_1 \setminus T_1} h^{(i)} {x_1 ^{(i)}}^{-\alpha}\)} > \theta\] \\ \nn
&= \mathsf{E}_{x,h} \[\exp\(-\frac{\theta  P_1d^\alpha h x^{-\alpha}}{P_2} \)\] \\ &\times \mathsf{E}_{\Phi_1, \{h^{(i)}\}} \[\exp(- \frac{\theta  P_1 d^\alpha \sum_{i\in \Phi_1 \setminus T_1} h^{(i)} {x_1 ^{(i)}}^{-\alpha}}{P_2} )\]. \label{prop1_def}
\end{align}\normalsize

Utilizing that the fading $h \sim \exp(1)$ provides 

\small\begin{align}
 &\mathsf{E}_{x,h} \[\exp\(-\frac{\theta  P_1d^\alpha x^{-\alpha}}{P_2} \)\]  =\mathsf{E}_x\[ \frac{P_2}{P_2 + {\theta  P_1d^\alpha x^{-\alpha}}}\] \\ 
 &=\frac{1}{2\pi} \int_0^\infty \frac{P_2}{P_2 + P_1 \theta d^\alpha \({R_I}^2 - 2 d {R_I} \cos(i) + d^2\)^{-\frac{\alpha}{2}}} di,
\end{align}\normalsize

\noindent where the last step follows from a triangular function when the angle between the primary TX $T_1$ and typical secondary RX has a uniform distribution.
Under the empty ball condition, we know that there is an empty ball of radius $R_I$ guaranteeing no primary TX inside. 
Imagine a thin circular ring $B_y$ of radius $y$ at a typical secondary RX. 
A secondary RX does not have any interferer at the intersected area of $B_y$ and empty ball. 
Then the intensity function of primary interferer becomes 

\small\begin{align}
\lambda'= \begin{cases} 
0 & {\textrm{if $ 0< y \leq \max(0,r-d)$,}} \\
2 \text{acos}\(\frac{{R_I}^2-d^2-y^2}{2 d y}\) \lambda_1  y&{\textrm{if $\max(0,{R_I}-d) < y \leq {R_I}+d$, }}  \\
2 \pi \lambda_1 y &{\textrm{if $R_I+d < y$.}}
\end{cases}\label{int_func}\normalsize
\end{align} \normalsize

By applying \eqref{int_func} and probability generating function of the PPP, we can represent the right part of \eqref{prop1_def} as 

\small\begin{align}\nn
&\mathsf{E}_{\Phi_1, \{h^{(i)}\}} \[\exp(- \frac{\theta  P_1 d^\alpha \sum_{i\in \Phi_1 \setminus T_1} h^{(i)} {x_1 ^{(i)}}^{-\alpha}}{P_2} )\] \\
\nn
&=\exp\left(-\lambda_1\int_{0}^{\infty}\frac{2\pi P_1 \theta  d^\alpha y^{-\alpha+1}}{P_2+P_1 \theta d^\alpha y^{-\alpha}}dy\) \\ \nn
&\ \times \exp\(\lambda_1\int_{|{R_I}-d|}^{{R_I}+d} \frac{2\text{acos}\left(\frac{{R_I}^2-d^2-y^2}{2 d y}\right)P_1 \theta d^\alpha y^{-\alpha+1}}{P_2 + P_1 \theta d^{\alpha} y^{-\alpha}} dy\) \\ 
&\ \times \exp\(-\lambda_1 \int_{\min(0,{R_I}-d)}^{{R_I}-d}\frac{2\pi P_1 \theta d^\alpha y^{-\alpha+1}}{P_2 +P_1 \theta d^\alpha y^{-\alpha}}dy\). \label{condsir_right}
\end{align} \normalsize

\subsection{Proof of Proposition 2}
In order to provide a tractable bound of the access probability, we modify the empty ball to have its center at the secondary RX location, with a radius of $\max(0,{R_I}-d)$.
Then the intensify function of primary interferer is 
\vskip -10pt
\small\begin{align}
\lambda' = \begin{cases}
0 &  {\textrm{if $ 0< y \leq \max(0,{R_I}-d)$,}} \\
2 \pi \lambda_1 y &{\textrm{if $\max(0,{R_I}-d) < y$.}} \end{cases}
\end{align}\normalsize

With this intensify function, the equation \eqref{condsir_right} becomes
\vskip -10pt
\small\begin{align}
&\exp\(-2\pi\lambda_1 \int_{\max(0,{R_I}-d)}^\infty \frac{ P_1 t d^\alpha y^{-\alpha+1}}{P_2 + P_1 t d^\alpha y^{-\alpha}} \) \\ 
&=\begin{cases} 
\exp\(-\pi\lambda_1 d^2 \(\frac{P_1 t}{P_2}\)^{\frac{2}{\alpha}} \int_{0}^\infty \frac{du}{1+u^{\frac{\alpha}{2}}}\) &{\textrm{if ${R_I}<d$,}} \\
\exp\(-\pi\lambda_1 d^2 \(\frac{P_1 t}{P_2}\)^{\frac{2}{\alpha}} \int_{\(\frac{d^\alpha P_1 t}{P_2  ({R_I}-d)^\alpha}\)^{-\frac{2}{\alpha}}}^\infty \frac{du}{1+u^{\frac{\alpha}{2}}}\) &{\textrm{if $d \leq {R_I}$.}} \end{cases} \label{eq_22}
\end{align}\normalsize

For the distance range $d \leq {R_I}$, we utilize the Pfaff transformation as Lemma 2 in \cite{JHpark}, $1+t^{\frac{2}{\alpha}}\int_{t^{-\frac{2}{\alpha}}}^\infty \frac{du}{1+u^{\frac{\alpha}{2}}} \leq \rho (1+t)^{\frac{2}{\alpha}}$.
Then the equation \eqref{eq_22} can be further simplified.


\subsection{Proof of Lemma 1}
Let $f_I(I)$ denote the distribution of the sensed interference and $f_{R|I}(R)$ the distribution of distance between a secondary TX and $T_1$ when $I$ is given. 
Then the average access probability becomes
\vskip -10pt
\small\begin{align}
\hat{\phi}_2(\theta) &= \int_0^\infty \int_0^\infty \mathcal{P}_s(R,\theta) f_{R|I} (R) dR f_I(I) dI \\&\overset{(a)}{=}  \int_0^\infty \mathcal{P}_s(r,\theta) f_r(r) dr,
\end{align}\normalsize

\noindent where $(a)$ follows from the change of integration order and $f_r(r)= 2\pi\lambda_1 r \exp(-\pi\lambda_1 r^2)$ from \cite{Andrews11}.

\subsection{Proof of Lemma 2}
Now that the signal experiences the Rayleigh fading, the transmission success probability can be divided as 
\small\begin{align}
\mathsf{P}\(\SIR_2 > \beta |I \) = \mathsf{P}\(\frac{h^{(0)} d^{-\alpha} P_2}{ \sum I_1} > \beta |I \)  \mathsf{P}\(\frac{h^{(0)} d^{-\alpha} P_2}{\sum I_2} > \beta |I \),
\end{align}\normalsize
where $\sum I_1$ denotes the sum of interference from primary TXs and $\sum I_2$ the sum of interference from secondary TXs, respectively.
The first probability is directly calculated from the access probability $\phi_2(I,\beta)$.
For deriving the second probability, we assume that transmitting secondary TXs are independently thinned with $\hat{\phi}_2(\theta)$.
Applying the mapping theorem \cite{Haenggi13} in $\SIR$ coverage \cite{Andrews11} completes the proof.
 
 \subsection{Proof of Lemma 3}
From the outage probability derived by using the Theorem 2 in \cite{Andrews11}, we can represent the outage probability of the primary network as 
\small\begin{align}\mathsf{P}\[\SIR_1\leq \gamma\] = 1-\frac{\lambda_1{P_2}^{\frac{2}{\alpha}}}{\lambda_2 \hat{\phi}_2\rho_0(\theta) {P_1}^{\frac{2}{\alpha}}+\lambda_1  {P_2}^{\frac{2}{\alpha}}( \rho(\theta) + 1) }. 
\end{align}\normalsize
Reminding the protection condition that $\mathsf{P}\[\SIR_1\leq \gamma\] < \tau$ finalizes the proof. 

\subsection{Proof of Proposition 3}

The ASE is represented as
\vskip -10pt
{\small\begin{align}
&\mathcal{A}= \lambda_2 \int_0^\infty  \phi_2(\theta,I) \mathsf{P}\(\SIR_2 > \beta |I\) \ln (1+\beta) f_I(I) dI \\
&= \lambda_2 \int_0^\infty \phi_2(I,\theta) \mathsf{P}\(\SIR_2>\beta | r\) \ln (1+\beta) f_r(r) dr, \label{ase_final}
\end{align} }\normalsize 

Combining the access probability and transmission success probability to derive \eqref{ase_final} and removing $\lambda_2$ that is independent on the control variable complete the proof.

\subsection{Proof of Proposition 4}
When $P_1 \lambda_1 \gg P_2 \lambda_2$, the interference from primary network is dominant over that from secondary network.
Thus, choosing the minimum access threshold $\theta$, $\overline{\theta}$, is optimal for increasing the number of transmission chance from the perspective of secondary network.
In addition, with the \textit{Leibniz integral rule}, we know that 
$
\frac{\partial}{\partial t}\int_a^b f(x,t) dx = \int_a^b \frac{\partial f}{\partial t} dx.
$
By utilizing this integral rule, we can differentiate the ASE with a access threshold $\overline{\theta}$ with respect to $\beta$, yielding the function $\mathsf{D}(\beta)$.


%

\end{document}